\documentclass[prb]{revtex4}  
\usepackage[]{graphicx}
\usepackage{latexsym,amssymb}

\def\mm#1#2#3#4{%
\left(
\begin{array}{lr}
\ds{#1}  &  \ds{#2}  \\
~~~   &  ~~\\
\ds{#3}  &  \ds{#4}
\end{array}
\right)
}

\def\mv#1#2{%
\left(
\begin{array}{l}
#1    \\
#2
\end{array}
\right)
}

\def\bea{\begin{eqnarray}}
\def\eea{\end{eqnarray}}
\def\be{\begin{equation}}
\def\ee{\end{equation}}
\def\ba{\begin{array}}
\def\ea{\end{array}}

\def\bc{\begin{center}}
\def\ec{\end{center}}
\def\ds{\displaystyle}  

\def\eps{\varepsilon}

\def\oA{\overline{A}}
\def\oB{\overline{B}}
\def\kmax{k_{\rm max}}
\def\zmax{z_{\rm max}}

\def\bS{\mathbf{S}}

\begin{document} 

\title{Propagation of surface plasmons through planar interface}

\author{Tom\'a\v{s} V\'ary  and Peter Marko\v{s}}
\affiliation{ Dept. Physics, Faculty of Electrical Engineering and Information Technology,\\
Slovak University of Technology, 812 19 Bratislava, Slovakia
}



\begin{abstract}
We analyze the scattering of the surface plasmon incident at a planar interface
between two dielectrics.
By using the scattering matrix technique, developed by Oulton et al. 
[Phys. Rev. B {\bf 76}, 035408 (2007)], we calculate 
the transmission, reflection coefficients and radiative losses  for oblique incident angles. 
We found that the transmission of a surface wave 
through a single interface between two dielectrics may be accompanied with radiation losses of 
10-40 per cent of the plasmon energy. 
\end{abstract}

  \maketitle 

\section{INTRODUCTION}
\label{sec:intro}  

The excitation of a surface electromagnetic wave at the metallic interface - surface plasmon
 \cite{economou,zayats} - opens new ways in 
nanophotonics \cite{krenn0,nano,prasad}  and metamaterial physics \cite{el-meta}.
One of the main constrain in  using of surface plasmons is their  short lifetime. 
It is well known \cite{lahm,stegeman0,stegeman,oulton,zayats} that a significant part of the 
plasmon energy is radiated  when the plasmon is scattered at the surface impurity.
A detailed quantitative analysis of the process of scattering and estimation
of radiation losses  is therefore  important for understanding of propagation
of surface plasmons.

The most simple scattering problem is the transmission and reflection of a surface plasmon at the
planar permittivity step, created when the metallic surface is covered by two  
different dielectrics \cite{stegeman,stegeman1,lahm,oulton}. In the most simple scattering
experiment, the metallic surface lies in the $z=0$ plane, and two dielectrics fill the
$z>0$ half-space. The interface between two dielectrics is given by $x=0$ plane, so that the
dielectric permittivity is
\be\label{eab}
\eps_d=\left\{
\begin{array}{ll}
\eps_a   &x<0\\
\eps_b   & x>0.
\end{array}
\right.
\ee
In this paper, we study the propagation  of the surface plasmon through  the planar interface 
between two dielectrics which cover the metallic surface.   
A modified method of Oulton \textsl{et al.} \cite{oulton}
is used for the calculation of the transmission and reflection
coefficient and analysis of   the radiation losses accompanying the plasmon scattering. 
Our data confirm that significant part of the surface plasmon  energy is radiated in the process of
single scattering, and energy  losses increase  when the angle of incidence increases.

\section{Surface plasmon at the metal - dielectric interface}

The surface plasmon propagates along
 the metal dielectric interface located in the $z=0$ plane. 
The intensity of the electric and magnetic field
 decays exponentially on both sides of the interface: $h\propto e^{-\kappa_d z}$ for $z>0$ (dielectric)
and $h\propto e^{+\kappa_mz}$ for $z<0$ (metal).  The parameters 
$\kappa_d$ and $\kappa_m$ are given by the 
dispersion relations,  \cite{economou,wp}
\be\label{spp-disp}
\ds{\frac{\kappa_m}{\kappa_d}} + \ds{\frac{\eps_m}{\eps_d}} = 0
\ee
($\eps_m$ is the metallic permittivity), and 
\be\label{spp-d}
\begin{array}{lcll}
k_\parallel^2 -\kappa_d^2 &=& k_0^2\eps_d  &  z>0\\
k_\parallel^2 -\kappa_m^2 &=& k_0^2\eps_m  &  z<0.
\end{array}
\ee
Here, $k_\parallel=\sqrt{k_x^2+k_y^2}$ is the projection of the wave vector into the $xy$ plane,
$k_0=\omega/c$ and $c$ is the light velocity.
From Eq. (\ref{spp-disp},\ref{spp-d}) we find explicit expressions for the components 
of the wave vector,
\be\label{spp-k}
k_\parallel^2 = k_0^2\ds{\frac{\eps_d\eps_m}{\eps_d+\eps_m}},~~~~
\kappa_d^2 = - k_0^2\ds{\frac{\eps_d^2}{\eps_d+\eps_m}},~~~~
\kappa_m^2 = - k_0^2\ds{\frac{\eps_m^2}{\eps_d+\eps_m}}.
\ee
These equations, together with the Drude expression for the metallic permittivity,
$\eps_m=1-\omega_p^2/\omega^2$, determines completely 
the frequency dependence of the wave vector of the surface plasmon.

The surface plasmon is TM polarized, with magnetic field parallel to the metal-dielectric 
interface.  The intensity of  magnetic and electric field is of the form
\be\label{spp-h}
h = {\cal{N}}_0(-\sin\theta,\cos\theta,0)~e^{i(k_xx+k_yy-\omega t)}\times
\left\{
\begin{array}{ll}
e^{-\kappa_dz}  &  z>0\\
e^{+\kappa_mz}  &  z<0,
\end{array}
\right.
\ee
and
\be\label{spp-e}
e = {\cal N}_0\ds{\frac{z_0}{k_0}} e^{i(k_xx+k_yy-\omega t)}
\times \left\{
\begin{array}{ll}
(+i\kappa_d\cos\theta, +i\kappa_d\sin\theta, -k_\parallel)e^{-\kappa_d z}/\eps_d & z>0\\
(-i\kappa_m\cos\theta, -i\kappa_m\sin\theta, -k_\parallel)e^{+\kappa_m z}/\eps_m  & z<0.
\end{array}
\right.
\ee
Here, $z_0=\sqrt{\mu_0/\eps_0}$ is the vacuum impedance, and $\theta$ determines the direction
of propagation in the $xy$ plane:
$\cos\theta=k_x/k_\parallel$, $\sin\theta=k_y/k_\parallel$.
Normalization constant  ${\cal N}_0$ is specified in Appendix \ref{sect-method}.

\subsection{Snell's law for surface plasmon}

\begin{figure}[t!]
   \begin{center}
   \includegraphics[width=7cm]{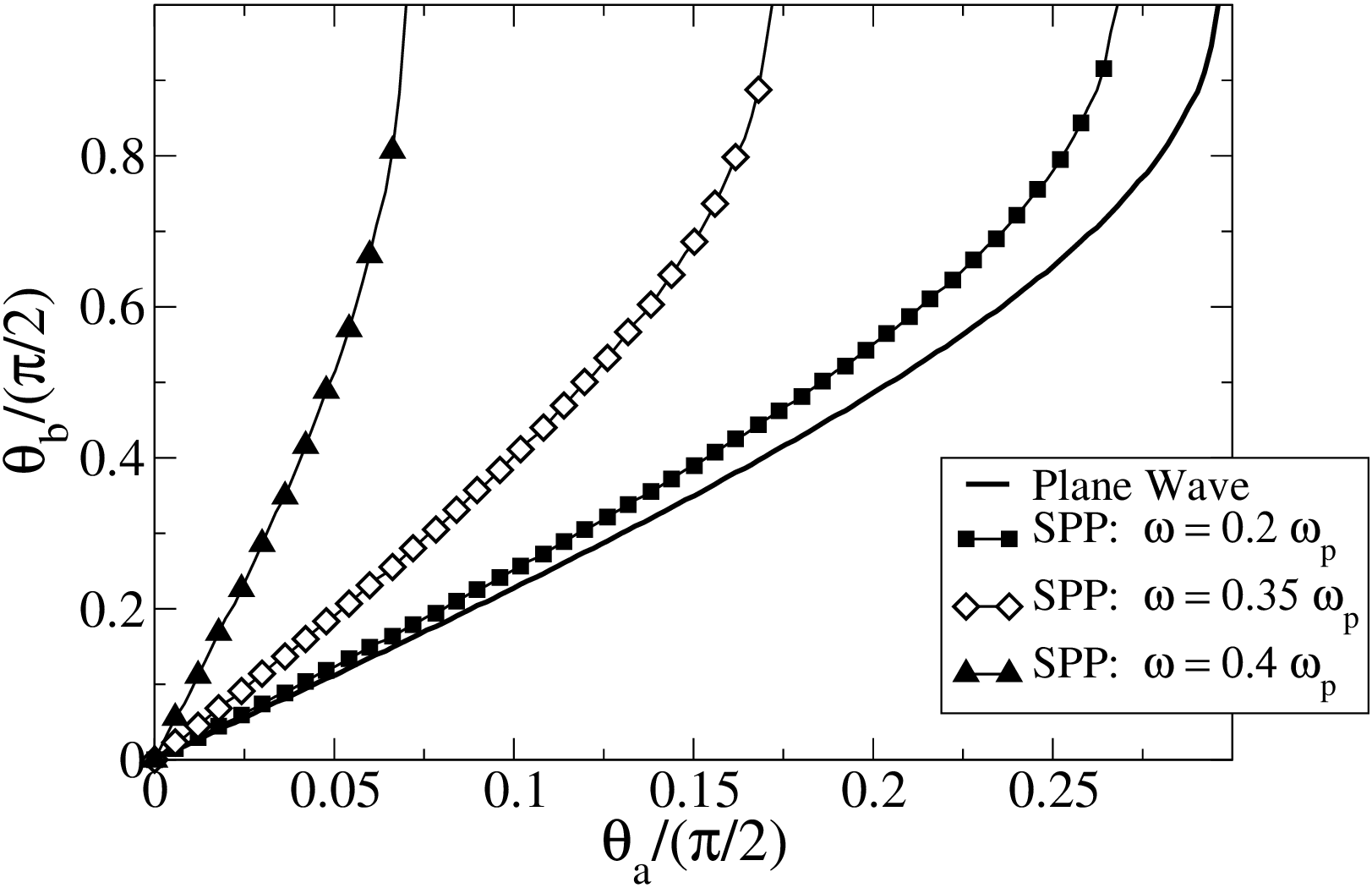}
   \end{center}
   \caption[example] 
   { \label{fig-spp-snell} 
Relation between two  angles $\theta_b$ and  $\theta_a$ for various frequencies  $\omega$ 
of the surface plasmon and 
permittivity step $\eps_b/\eps_a = 5$. In the case of  the transmission $a\to b$,  the 
reflection angle $\theta_b$ is always larger  than that for the plane wave.
Consequently, in the case of the transmission $b\to a$ the critical angle $\theta_{cP}$ 
is always smaller than the critical angle $\theta_c$ for plane waves.
}
   \end{figure}

In the scattering experiment, we consider the    metal  covered by two different dielectrics, 
$a$ and $b$, with permittivities $\eps_{a}$ and $\eps_{b}$.
The interface between dielectrics lies in the $yz$ plane $x=0$. 
From the  continuity of the 
$y-$ component of the wave vector, $k_{ya}=k_{yb}$,  we find the relation between the incident
and the refractive angle,
\be\label{spp-snell}
\ds{\frac{\sin\theta_b}{\sin\theta_a}}=\ds{\sqrt{\frac{\eps_a}{\eps_b}}}\ds{\sqrt{\ds{\frac{\eps_b+\eps_m}{\eps_a+\eps_m}}}}.
\ee
Figure \ref{fig-spp-snell} shows $\theta_b$ as a function of $\theta_a$ for various values of  
the plasmon frequency $\omega$.  The most important consequence of the relation (\ref{spp-snell}) 
is the existence of
the critical angle $\theta_{cP}$ for the surface plasmon incident from the media
with higher permittivity.  No transmission of surface plasmon is
 possible when the incident angle $\theta_b>\theta_{cP}$.

\section{Transmission and reflection coefficients for the surface plasmon}

The transmission and reflection coefficients will be calculated from the requirement of the
continuity of tangential components of both electric and magnetic fields at
 the interface $x=0$ between two dielectrics $a$ and $b$,
\be\label{cont1}
E_{za}(x\to 0^-)\equiv E_{zb}(x\to 0^+)~~~~\textrm{and}~~~~ 
H_{ya}(x\to 0^-)\equiv H_{yb}(x\to 0^+).
\ee
Since  the intensity of the surface plasmon decreases exponentially in   the $z$ direction 
(Eqs. \ref{spp-h},\ref{spp-e}), two plasmons on the opposite sides of the interface cannot satisfy 
the continuity relations (\ref{cont1})  for all $z$. Therefore,
we have to consider the full system of eigenwaves for the metal-dielectric interface. 
This system contains, besides the
surface plasmon,  an infinite number of plane waves.  All plane wave
have the same $y-$ component of the wave vector, but differ  in the $z-$ component $k_z$.
In our numerical  analysis, we use   $N$ plane waves 
with  $k_{z\alpha} = (\alpha/N)\kmax$ and $\alpha=1,2,\dots ,N$. The upper cutoff $\kmax$ is 
specified in Appendix  \ref{sect-method}.

With the use of the plane waves, we express the  continuity equations for the 
electric and magnetic field given by Eq. (\ref{cont1}) in the form
\be\label{eq-1}
\sum_{\alpha=0}^N (A_\alpha-\oA_\alpha)E_{za\alpha} = 
\sum_{\alpha=0}^N (B_\alpha-\oB_\alpha)E_{zb\alpha} 
\ee
for the $z$-components of the electric field $E_z$, and
\be\label{eq-2}
\sum_{\beta=0}^N (A_\beta+\oA_\beta)H_{ya\beta} = 
\sum_{\beta=0}^N (B_\beta+\oB_\beta)H_{yb\beta}
\ee
for the $y$ components of the magnetic field $H_y$. Fields $E_z$ and $H_y$ are given in the
Appendix \ref{sect-method}.
Vectors  $A_\alpha = A(k_{z\alpha})$ 
and $B_\beta=B(k_{z\beta})$,  $\alpha, \beta = 0,1,\dots ,N$
contain  amplitudes of the surface plasmon ($\alpha,\beta=0$) and  
 $N$  plane waves in the left  and right media, respectively. 
$A$, $B$  ($\oA$, $\oB$) represent fields propagating to the right and to the left,
respectively.

All amplitudes  can be calculated from the requirement that
Eq.  (\ref{cont1}) must be fulfilled for any  $z$ \cite{stegeman0,stegeman}. 
Another approach, which uses the coupling coefficients between electric and
magnetic fields, suggested in Ref. \cite{oulton},
enables us to formulate  the problem in terms
of the  $2(N+1)\times2(N+1)$ scattering matrix $\bS$
\be\label{sm}
\mv{B}{\oA}=\mm{\bS^{bb}}{\bS^{ba}}{\bS^{ab}}{\bS^{aa}}\mv{\oB}{A},
\ee
which relates the amplitudes of the incoming waves $A$ and $\oB$ with the
outgoing waves $\oA$ and $B$.
Details of the calculation are given in Appendix \ref{sect-method}. 

In the next Section, we analyze  the  case when the only incident wave is the surface plasmon 
propagating in media $a$.
Then $A_\alpha = \delta_{0\alpha}$ and $\oB\equiv 0$. 
From Eq. (\ref{sm}) we obtain the transmission 
and reflection coefficients for the surface plasmon, 
\be
T_{a\to b} = |\bS^{ba}_{00}|^2~~~~\textrm{and}~~~ 
R_{a\to a} = |\bS^{aa}_{00}|^2.
\ee
The components $\bS^{aa}_{\alpha0}$ and 
$\bS^{ba}_{\beta 0}$  determine radiation losses due to the scattering of the surface plasmon.
Among all plane waves, only those with real $k_x$  radiate the energy in the $x$ direction.
Since $k_{ax\alpha}^2=k_0^2\eps_a - k_{y}^2-k_{az\alpha}^2$, we have that $k_{ax\alpha}$ 
is real only  
for $\alpha$ smaller than certain integer $n_a$. Similarly,   $k_{bx\beta}$ is real only when
$\beta<n_b$.
Total radiation losses are therefore obtained as $S_a=S_{aa}+S_{ba}$, where
\be
S_{aa} = \sum_{\alpha}^{n_a} |\bS^{aa}_{\alpha 0}|^2
~~~\textrm{and}~~~ 
S_{ba} = \sum_{\beta}^{n_b} |\bS^{ba}_{\beta 0}|^2.
\ee
The conservation of the energy  requires
\be
T_{a\to b}+R_{a\to a} + S_a = 1.
\ee
Physical meaning of other components of the scattering matrix is obvious. 
For instance, the element
$\bS^{ab}_{0\beta}$ gives the amplitude of a surface plasmon, 
excited in the media $a$ by a plane wave $\beta$
incident to the interface from the media $b$. 

\subsection{Normal incidence}

   \begin{figure}[t!]
   \begin{center}
   \includegraphics[width=7cm]{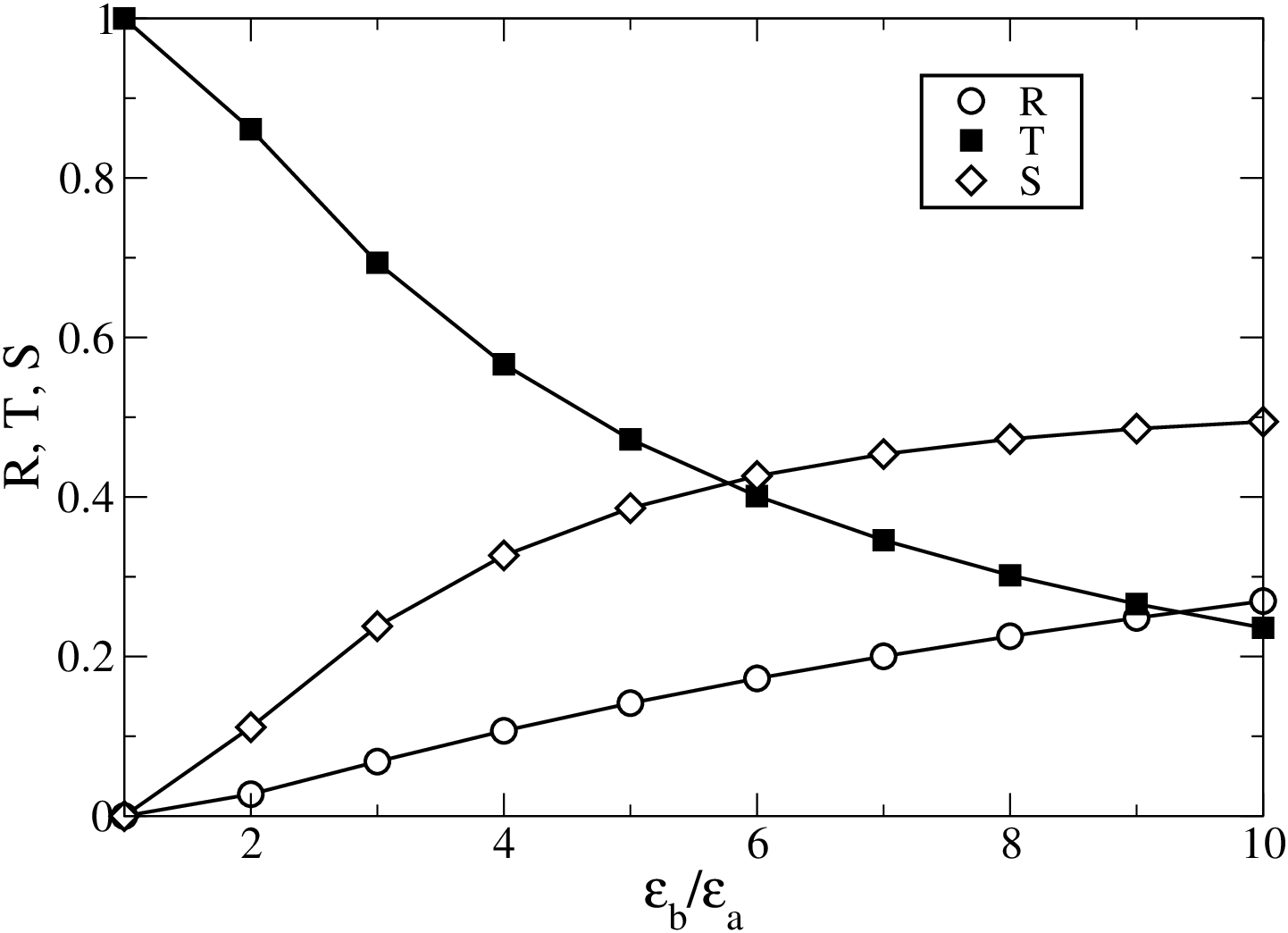}~~~~~~
   \includegraphics[width=7cm]{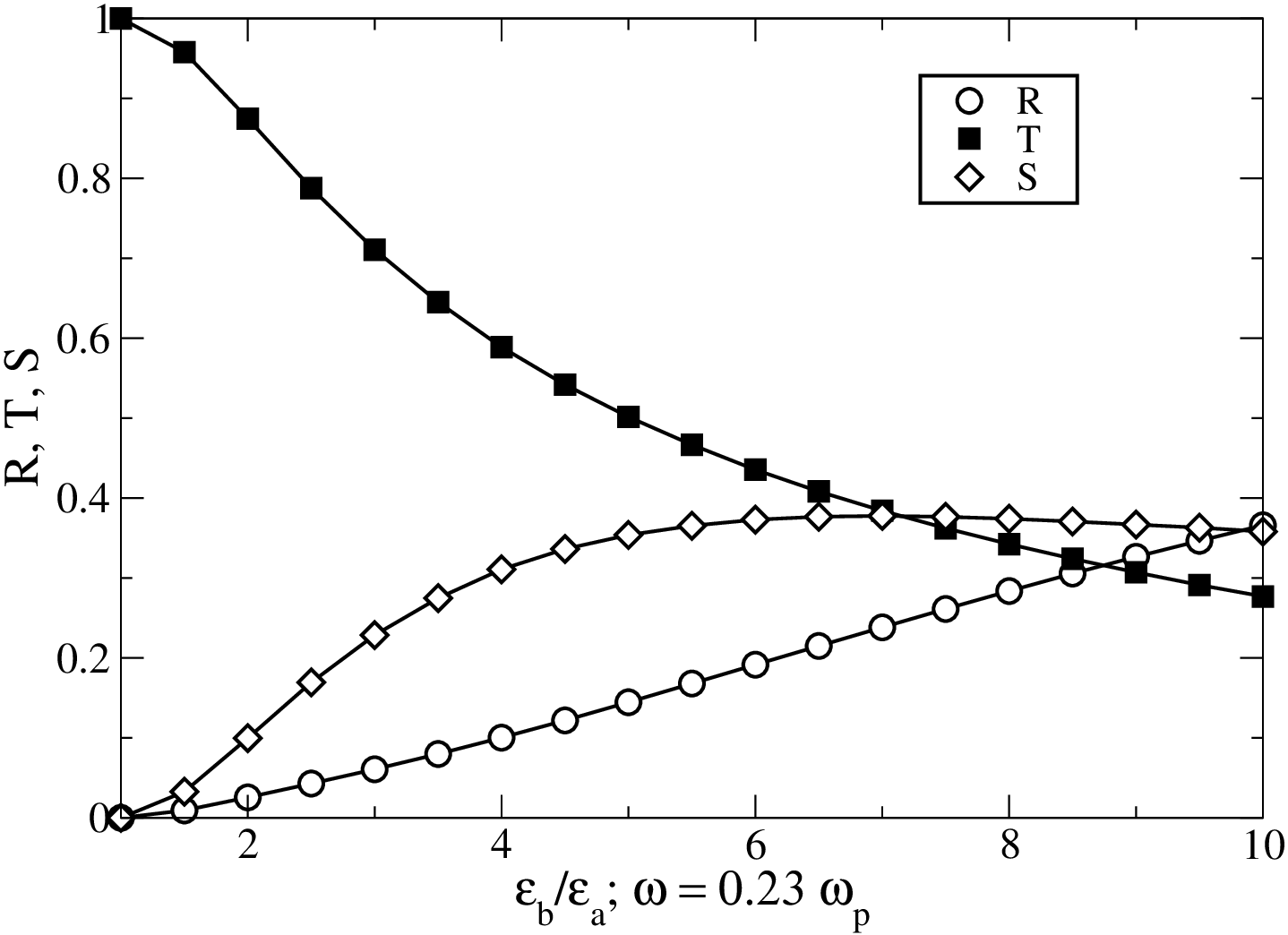}
   \end{center}
   \caption[example] 
   { \label{fig-1to2} 
Transmission $T$, reflection $R$ and radiative losses $S$  for the normal incidence of the   
surface plasmon
at the interface between two dielectrics.
The surface plasmon is coming from the dielectric $a$ (left) and from the dielectric $b$ (right).
}
   \end{figure} 

\begin{figure}[b!]
   \vspace*{0.7cm}
   \begin{center}
   \includegraphics[width=4cm]{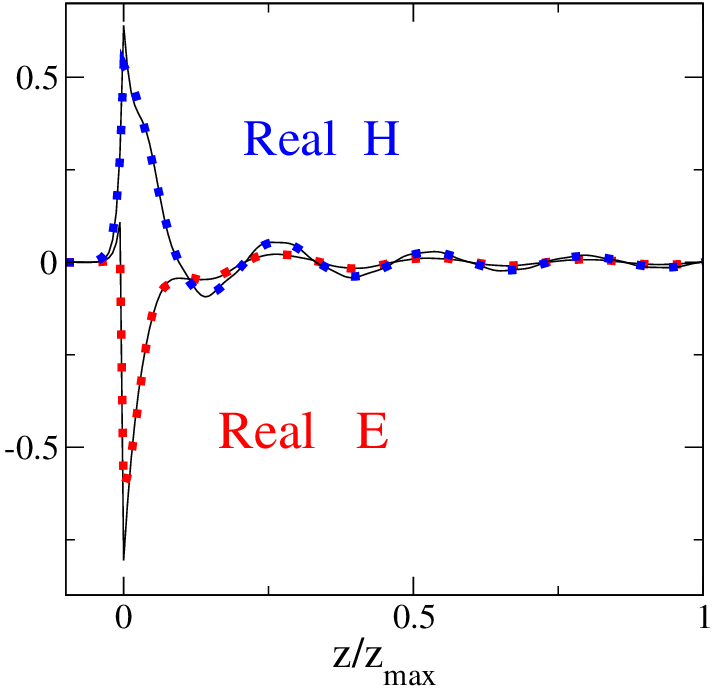}~~~~~~~~
   \includegraphics[width=4cm]{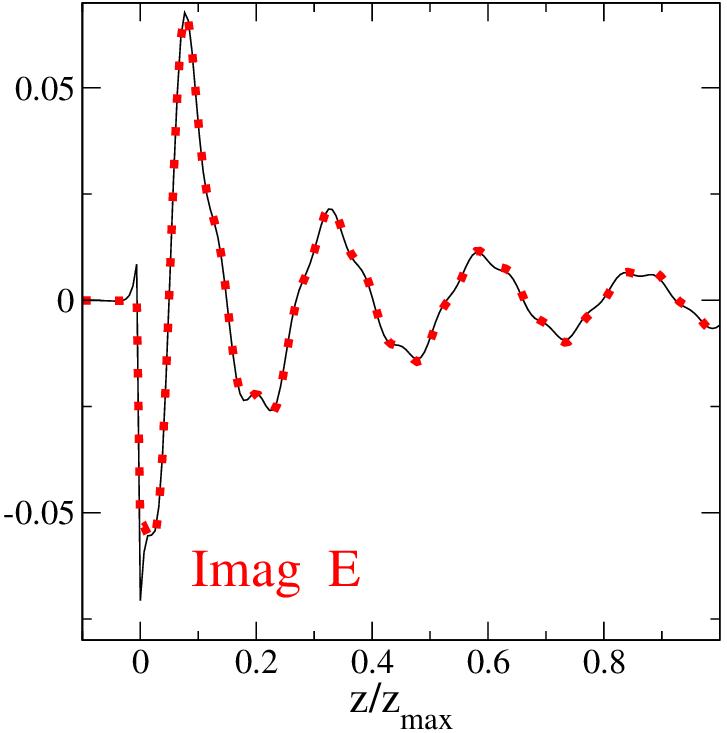}~~~~~~~
   \includegraphics[width=4cm]{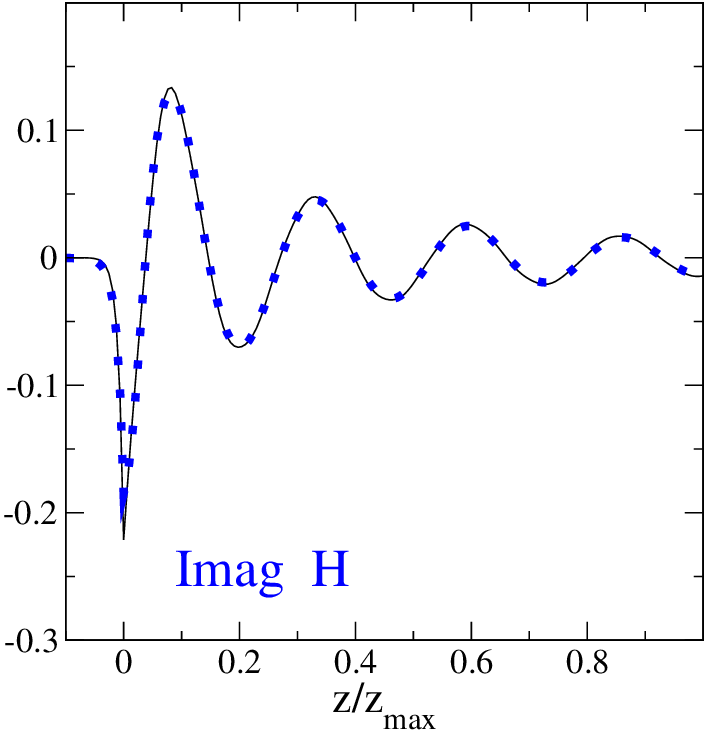}
   \end{center}
   \caption[example] 
   { \label{fig-polia} 
Test of the continuity of electric and magnetic field along the interface $x=0$. Solid lines 
and symbols represent fields for $x\to 0^-$ and $x\to 0^+$, respectively. Left figure show
real part of both $E$ and $H$, and two other figures show imaginary parts of fields. 
Dielectric permittivities are 
$\eps_a=1$, $\eps_b=5$.  $N=1577$ plane waves were used with with maximal $z-$ component
of wave vector given by Eq. (\ref{kmax}). The  maximal value of $z$ is given as
$\zmax=32/\kappa_{da}$.
}
   \end{figure}

\begin{figure}[t!]
   \begin{center}
   \includegraphics[width=7cm]{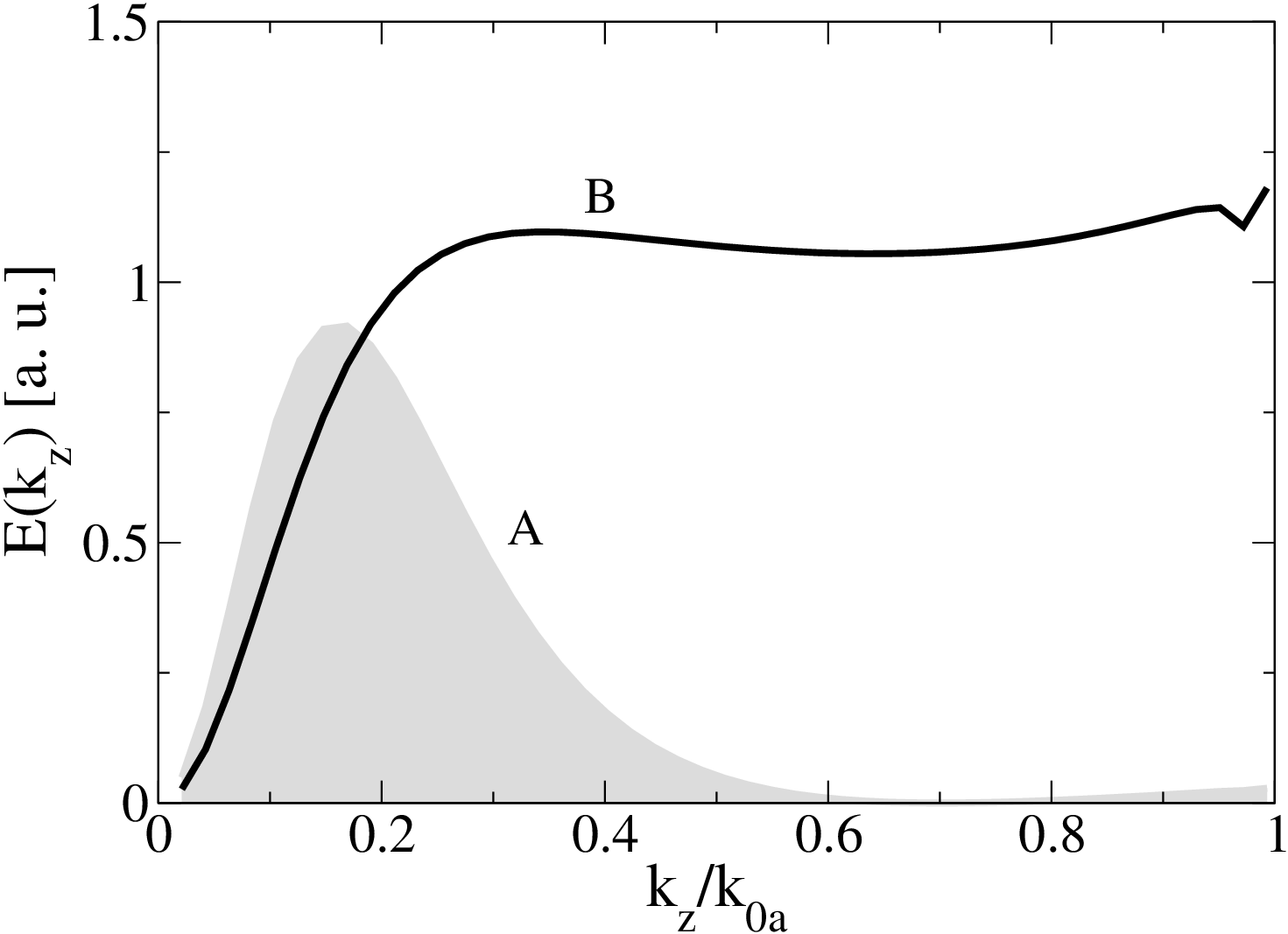}~~~~~~~~~~~~~
   \includegraphics[width=7cm]{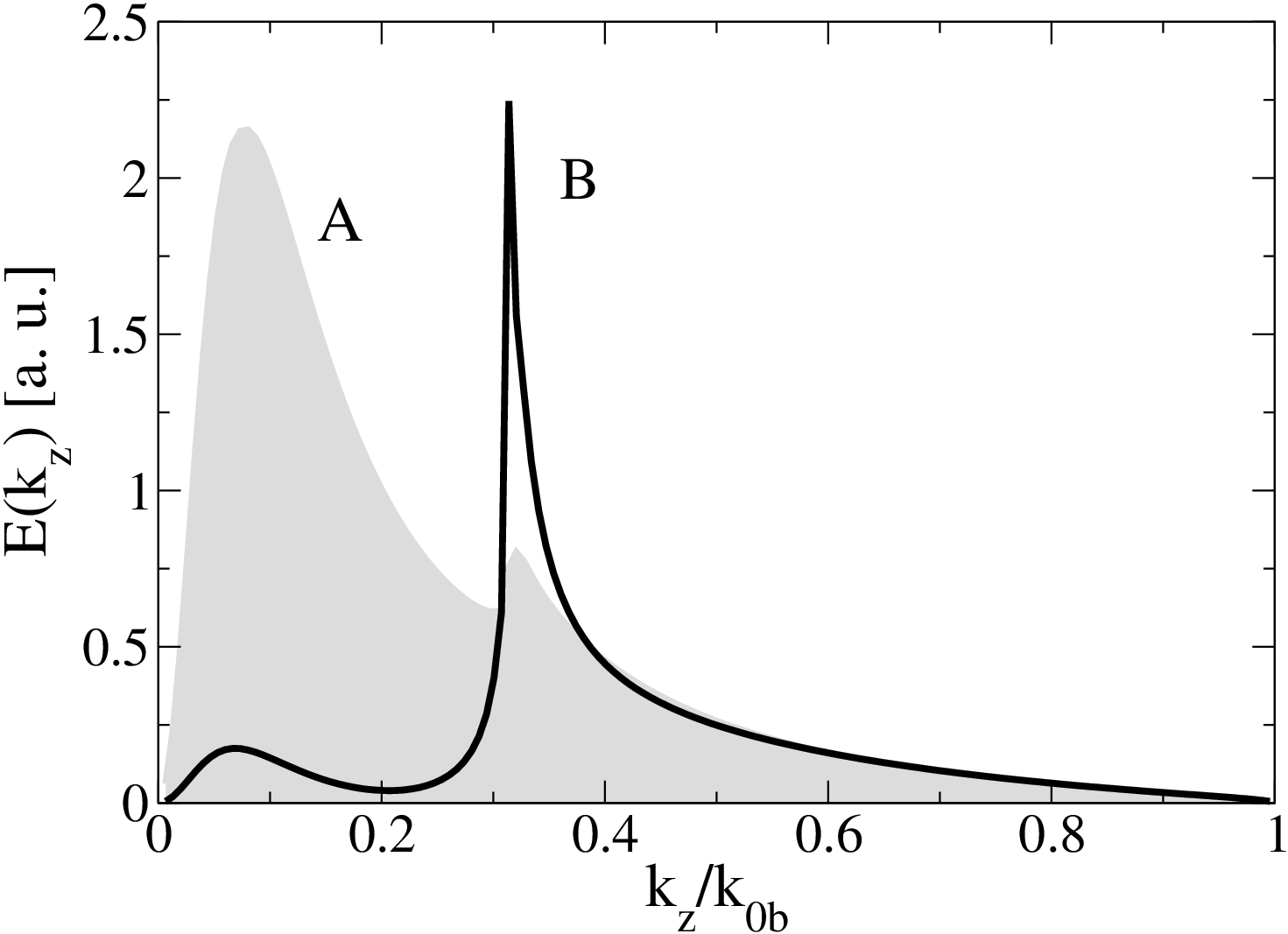}
   \end{center}
   \caption[example] 
   { \label{fig-uhlove} 
Angle distribution of radiative losses $E(k_z)$ (in arbitrary units) given by scattering matrix
elements $\bS^{ab}_{\alpha 0}$ for the scattering of the surface plasmon 
with the frequency $\omega=0.23\omega_p$
scattered at the interface between two dielectrics with permittivities
$\eps_a=1$, $\eps_b=10$.  $k_{0a}=k_0\sqrt{\eps_a}$ and $k_{0b}=k_0\sqrt{\eps_b}$.
Left (right)  figure shows radiation in the medium $a$ ($b$), respectively. $A$ and $B$
determines the medium from which plasmon is coming. 
Sharp maximum in the right figure shows the scattering in the direction
of the critical angle $\theta_c$ for plane waves.
}
   \end{figure}

Figures \ref{fig-1to2} - \ref{fig-uhlove} show scattering parameters  of the
surface plasmon for the case of normal incidence. In numerical calculations, we use
$\kmax$ given by Eq. (\ref{kmax})  and  number of plane waves varies between 
$N=100$ and  $N= 1577$.


The transmission and reflection coefficients as well as radiative losses 
are given in 
Fig. \ref{fig-1to2}  for various values of the
permittivity steps  $\eps_b/\eps_a$.  
Our data agree with  results of  Ref. \cite{oulton} and confirm   that
a significant part of the plasmon energy is radiated by plane waves. 
Radiation losses increase when the  permittivity step increases.  On the other hand,
scattering coefficients depend only weakly  on   the plasmon 
frequency (data not shown).  

As the test of numerical accuracy of the method, we used  amplitudes $\oA$ and $B$,
obtained from the scattering matrix, 
and  reconstruct  the electric and magnetic fields on both sides
of the $x=0$ plane for the permittivity step $\eps_b/\eps_a=5$.
Figure  \ref{fig-polia} confirm that  the tangential components of $E_z$ and
$H_y$ are indeed continuous at the interface.

In Fig. \ref{fig-uhlove} we present the  amplitudes of radiated plane waves, 
$|\bS^{ab}_{\alpha 0}|^2$
for the  plasmon incident from the media $a$ and $b$. The data  confirm that 
the energy is mostly radiated in the direction of incoming plasmon. 
As shown in the right figure,  the radiation 
possesses the sharp maximum in the direction of the critical angle for planar waves $\theta_c$.

   \begin{figure}[b!]
\vspace*{1cm}
   \begin{center}
   \begin{tabular}{c}
   \includegraphics[width=7cm]{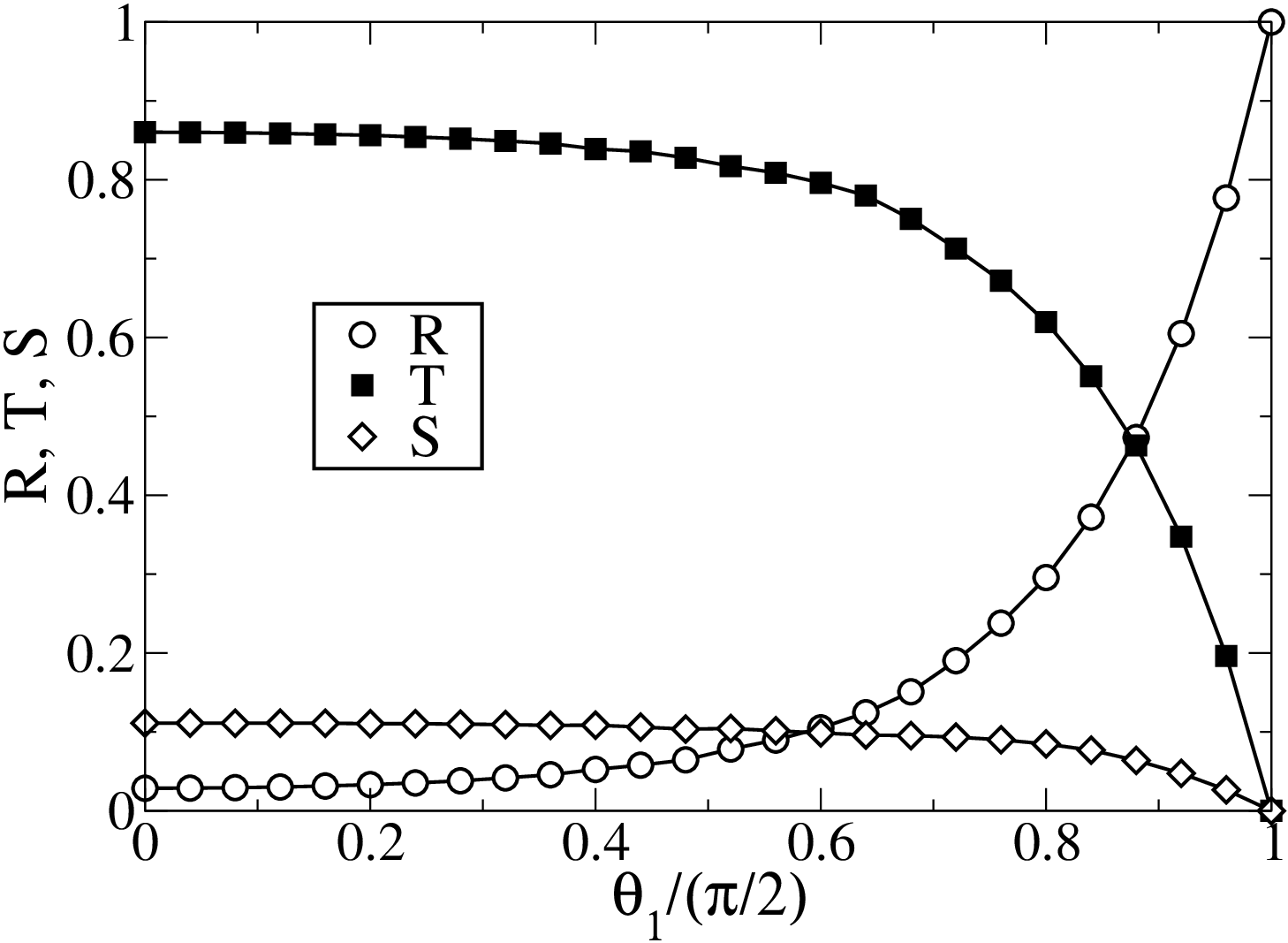}~~~~~~~~~~~
   \includegraphics[width=7cm]{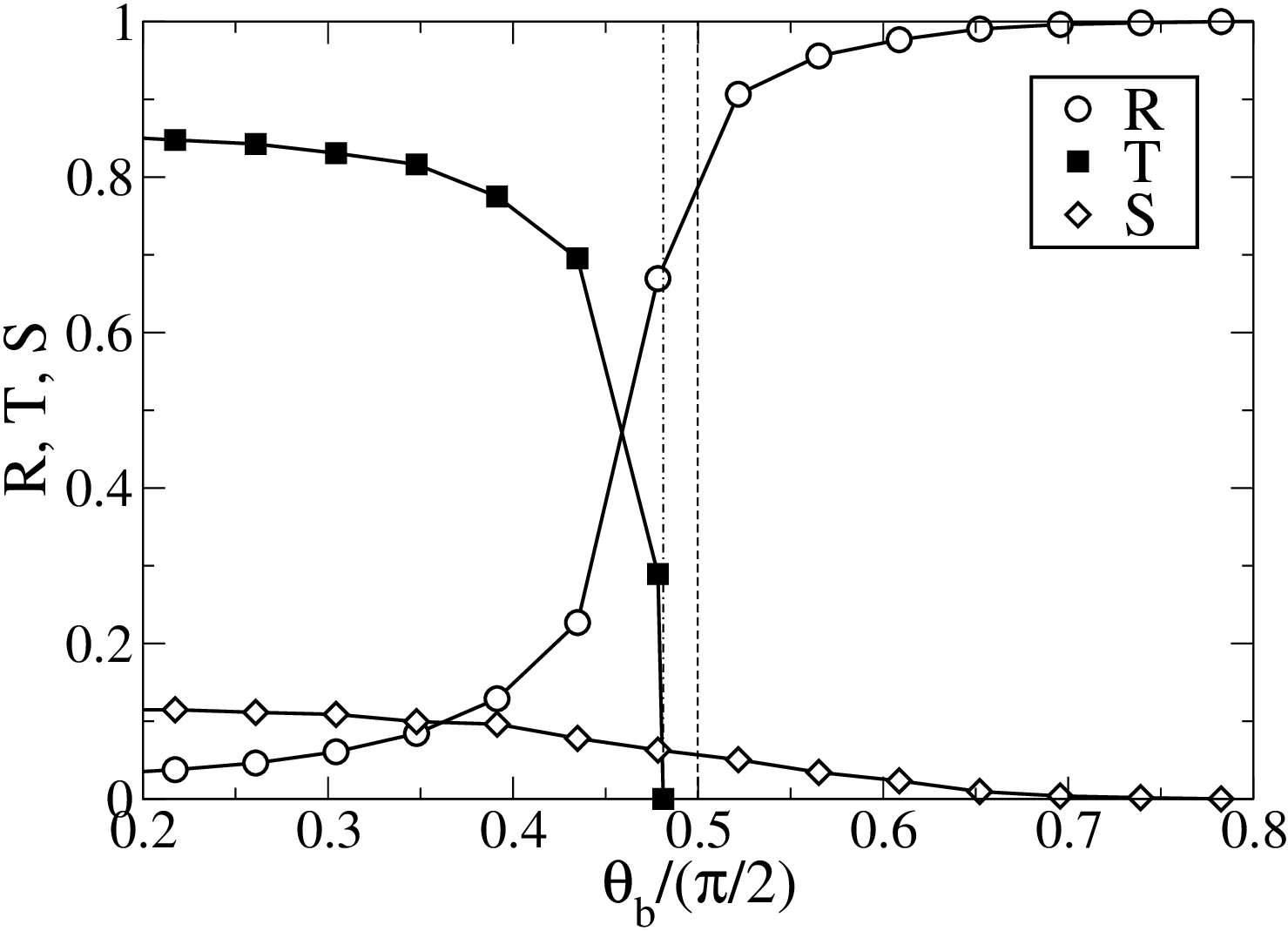}
   \end{tabular}
   \end{center}
   \caption[example] 
   { \label{fig-thetaEps2om023} 
Transmission $T$, reflection $R$ and radiative losses $S$  as a function of incidence angle  for
the
frequency $\omega = 0.23\omega_p$ and dielectric interface $\eps_a=1$ and $\eps_b=2$.
Left: scattering $a\to b$. Right: scattering $b\to a$. 
Vertical dashed lines show critical angles
 $\theta_{cP}$ for  the surface plasmon and $\theta_c$ for plane waves.
}
   \end{figure} 

\subsection{Oblique incident angle}

\begin{figure}[t!]
   \begin{center}
   \includegraphics[width=7cm]{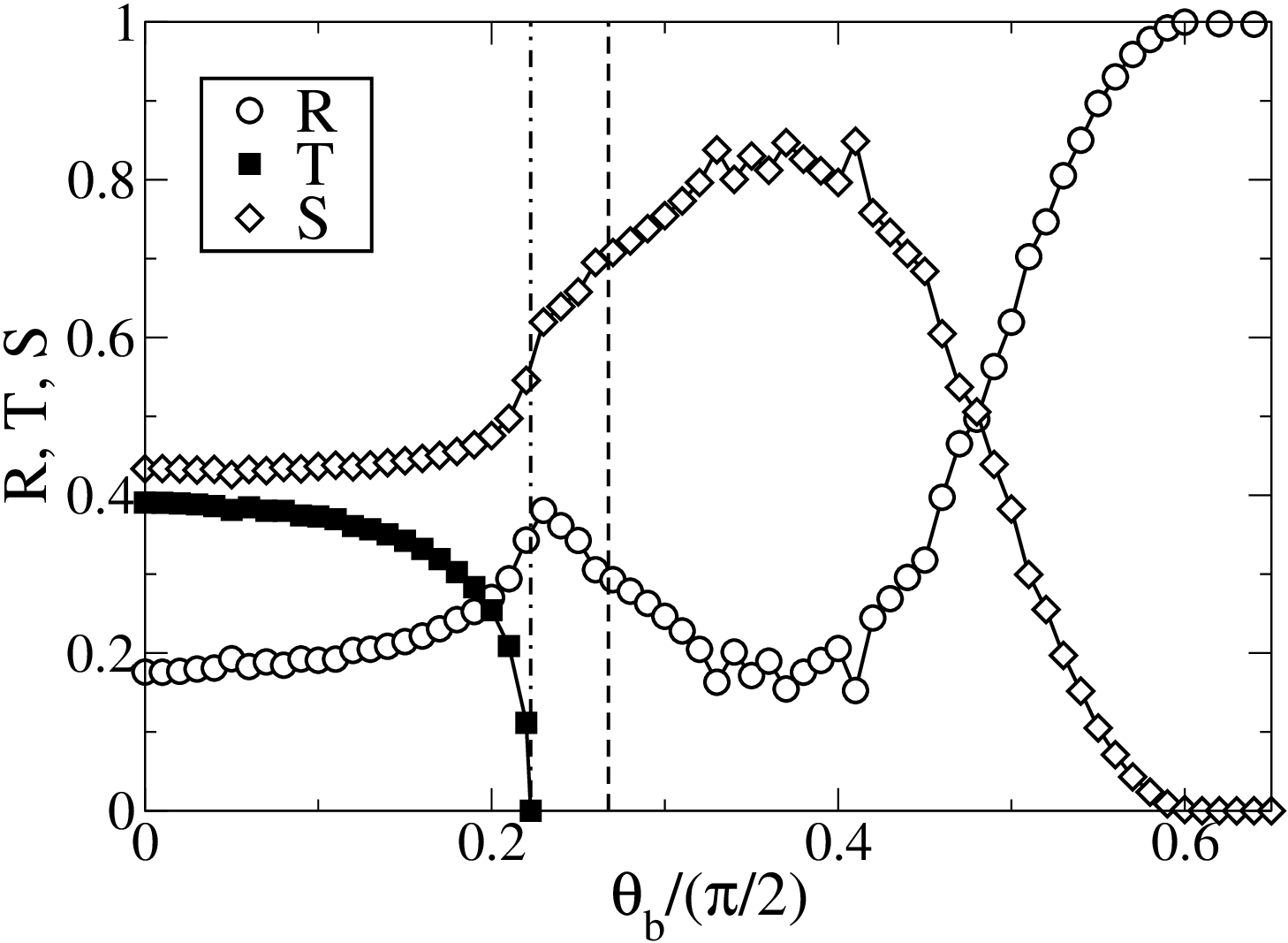}~~~~~~~~~
   \includegraphics[width=7cm]{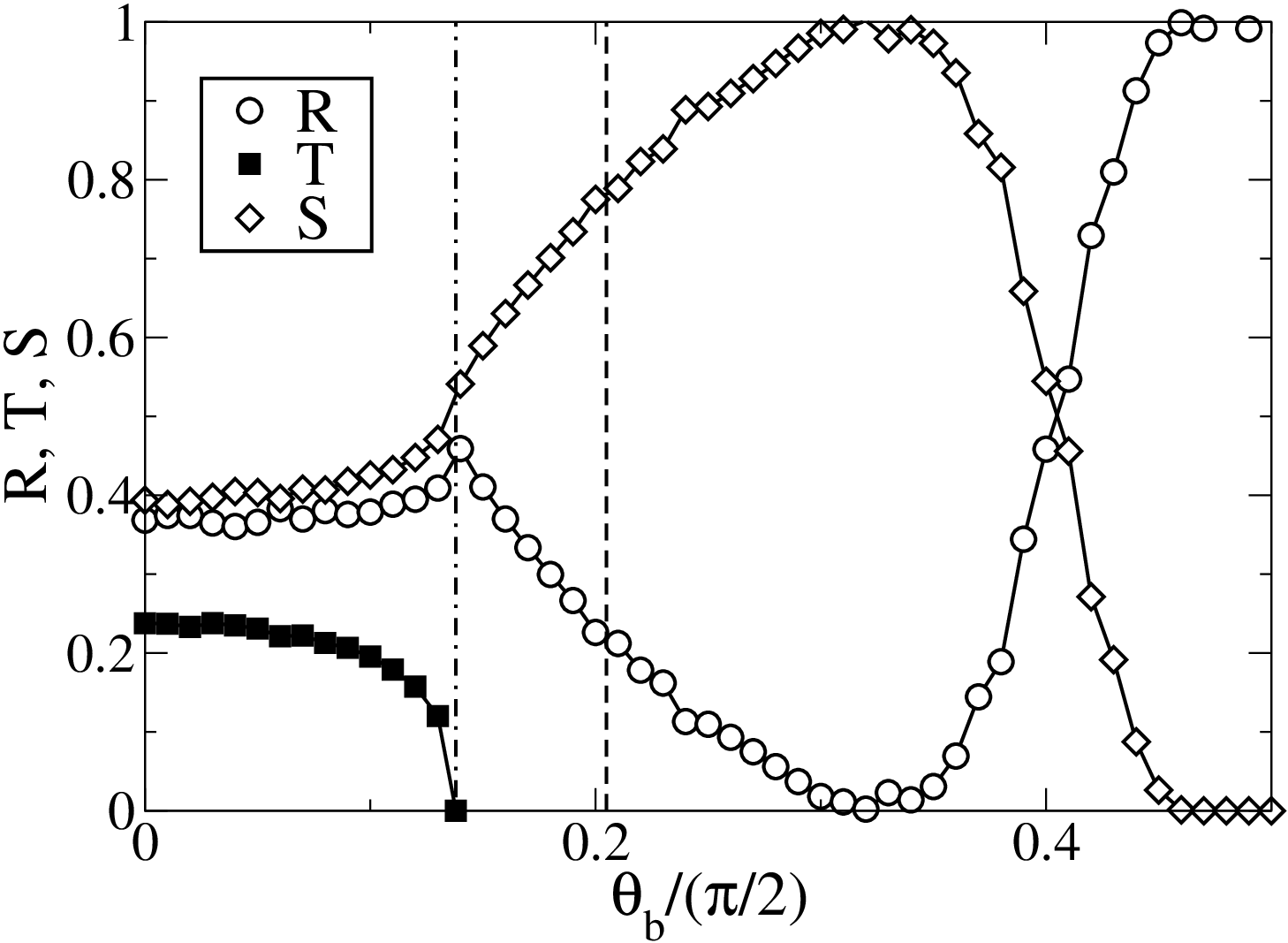}
   \end{center}
   \caption[example] 
   { \label{fig-thetaEps6om023b} 
Transmission $T_{b\to a}$, reflection $R_{b\to b}$ and radiative losses $S_b$  as a function of incidence angle  for
the
frequency $\omega = 0.23\omega_p$ and interface $\eps_a =1 $ and $\eps_b=6$  (left)
and $\eps_b=10$ (right).
Vertical dashed lines show critical angles
 $\theta_{cP}$ for  the surface plasmon and $\theta_c$ for plane waves. The transmission is zero
when $\theta>\theta_{cP}$.
The reflection decreases for $\theta>\theta_{cP}$ in favor of the radiation losses $S_b$. 
Total reflection is observable only for angles
$\theta\gg \theta_c$.
}
   \end{figure}

Figures \ref{fig-thetaEps2om023}  and \ref{fig-thetaEps6om023b} 
show how the scattering coefficients depend on the 
incident angle.  The plasmon is approaching the permittivity step either 
from the left or from the right side  of the interface.

For the  scattering from the side with lower permittivity, $\eps_a<\eps_b$,
all coefficients   depend monotonously on the incident angle.
More interesting is the case when the plasmon approaches the interface 
from the side with higher permittivity $\eps_b$.
The transmission, $T_{b\to a}$ 
decreases to zero when $\theta\to \theta_{cP}$, but the reflection $R$
does not increase to the  unity.  We explain this behavior by the presence of
``evanescent plasmon'' in the media $a$. Although the $x-$ component of the plasmon
wave vector $k_{ax}$ is imaginary, the intensity of the field on the left side of the interface
is non-zero (even larger than for smaller incident angles). This field must be  compensated
by plane waves which radiate energy.

For higher permittivity contrast $\eps_b/\eps_a$, we found that the reflection even 
decreases when the incident angle increases above the critical angle
$\theta_{cP}$. This decrease is accompanied by higher radiation losses. 
As shown in Fig. \ref{fig-thetaEps6om023b} 
entire  plasmon energy can be radiated when $\theta> \theta_{cP}$. 
Radiation losses have a maximum for the incident angle larger than the critical angle.
The ``total reflection'' ($R=1$)   takes place only for angles 
much larger than the critical angle for the surface plasmon.

\section{Conclusion}

We analyzed quantitatively the scattering of the surface plasmon at 
the planar interface between two
dielectrics which cover the metallic surface.  The transmission, reflection coefficients and
radiative losses were calculated for the normal and oblique incident angle.
We confirm that  the radiation of plane waves
causes significant scattering losses: for normal incidence, 
the transmission through the single interface might 
cost 20-40 \% of the plasmon energy.  The
 reduction of these losses represent the challenging problem for the theoretical 
research.
One possible way how to avoid this problem is to cover the metallic surface by 
anisotropic metamaterial instead of a dielectric \cite{meta}. 

We analyzed how the transmission and reflection coefficients depend on the incident 
angle. While the 
transmission to the dielectrics with $\eps_b>\eps_a$ brings no surprising result, the transmission in the
opposite direction exhibits non-monotonous dependence on the incident angle.  
The transmission coefficient 
decreases to zero when the incident angle increases to the  critical angle $\theta_{cP}$ for the
 surface plasmon. However, the reflection does not reach unity for the critical angle, because
the significant part of the energy is radiated. 

In the present analysis, we used real (losless) metallic permittivity. This is consistent with the
formulation of the scattering experiment, in which the incident wave is coming from the infinity.
Nevertheless,  we verified that   realistic losses, given by 
small imaginary part of the metallic permittivity in the Drude formula, 
 do  not influences the scattering coefficients.

\appendix    

\section{The method}
\label{sect-method}

Since the single  surface plasmon cannot satisfy the continuity relations (\ref{cont1}),
along the dielectric interface,
a complete set of plane waves must be included into the scattering procedure. 
Each plane wave is given by  a superposition of the  wave incident to
 ($\propto e^{-ik_{dz}z}$),
 reflected from 
($\propto re^{+ik_{dz}z}$) and
transmitted  through   
the metal-dielectric interface ($\propto te^{-ik_{mz}z}$). Here $k_{dz}$ and $k_{mz}$
are the $z-$ components of the wave vector in dielectric and metal,
respectively. The reflection amplitude $r$  for the metal-dielectric interface is given by
\be
r=\ds{\frac{\eps_dk_{mz}-\eps_mk_{dz}}{\eps_{d}k_{mz}+\eps_{m}k_{dz}}}
\ee
and $t=1-r$. In numerical calculation,
we  consider   $N$ plane waves 
with different values of the $z$-component of the wave vector
\be
k_{dz\alpha} = \ds{\frac{\kmax}{N}}\times \alpha, ~~~~\alpha=1,2,\dots ,N.
\ee
Here, $\kmax$ is the largest allowed value of $k_z$. We choose
\be\label{kmax}
\kmax = k_0\sqrt{\eps_d-\eps_m}
\ee
where $\eps_d=\textrm{min}(\eps_a,\eps_b)$. This choice guarantees that all waves transmitted
from the dielectric to the metal decrease exponentially, so that 
no plane wave propagates  inside  the metal and the $z-$ component of the wave vector in the metal,
\be\label{kmz}
k_{zm}= \sqrt{k_{dz}^2-\kmax^2},
\ee 
is imaginary.

In what follows we need explicit form of the 
the plane wave  for the interface metal-dielectric $a$ 
the fields $E_z$ and $H_y$.  
Neglecting the phase factor $\exp[i(k_xx+k_yy-\omega t)]$, we have
\be
H_{ya\alpha}(\vec{r}) = {\cal N}_{a\alpha}\ds{\frac{k_{ax\alpha}}{k_{a\parallel}}} 
\times
\left\{
\begin{array}{ll}
\left[ -e^{-ik_{az\alpha}z}+ r_{a\alpha} e^{ik_az\alpha}z\right]   & z>0\\
~~~\left[ - t_{a\alpha} e^{-ik_{mza\alpha }z}\right]   & z<0,
\end{array}
\right.
\ee
and
\be
E_{za\alpha}(\vec{r}) = - {\cal N}_{a\alpha} k_{a\parallel} \ds{\frac{z_0}{k_0}} 
\times
\left\{
\begin{array}{ll}
\left[ -e^{-ik_{az\alpha}z}+ r_{a\alpha} e^{ik_{az\alpha}z}\right]/\eps_d   & z>0\\
~~~\left[ - t_{a\alpha} e^{-ik_{mza\alpha}z}\right]/\eps_m   & z<0
\end{array}
\right.
\ee
with $k_{mza\alpha}$ given by Eq. (\ref{kmz}).
Similar expression can be written for the interface metal-dielectric $b$.

To solve the system (\ref{eq-1},\ref{eq-2}), we  multiply  both sides 
of Eq. (\ref{eq-1}) by  $H_{ya\alpha}$
and  Eq. (\ref{eq-2}) by $E_{zb\beta}$ and integrate over  $z$. Since all waves
possess the same the $y-$ component of the wave vector,
the coupling coefficient between two plane waves is given by the integral 
 \cite{schevchenko,oulton} 
\be\label{cab}
C_{\alpha\beta}^{ab} = \int_{-\xi}^{+\xi} dz E_{za\alpha} H_{yb\beta}, 
\ee
where $\xi$ is infinity when $a\ne b$ and $\xi=\zmax=(2\pi/\kmax)N$ for $a=b$. 
The choice of the spatial cutoff $\zmax$ 
enables us to express all integrals (\ref{cab}) in terms of dimensionless parameters.
We find that diagonal elements  of the matrix $C$ reads
\be
C_{\alpha\alpha}^{ab} = -{\cal N}_{a\alpha}{\cal N}_{b\alpha}\ds{\frac{z_0}{k_0}\frac{k_{a\parallel}}{k_{b\parallel}}\frac{\pi N k_{ax}}{\kmax}}(r_{a\alpha}+r_{b\alpha}).
\ee
The requirement $C^{aa}_{\alpha\beta}=\delta_{\alpha\beta}$ determines the norm 
${\cal N}_{a\alpha}$:
\be\label{na}
{\cal N}_{a\alpha} = i\ds{\sqrt{\ds{\frac{k_0}{z_0}\frac{\kmax}{2\pi N}\frac{\eps_a}{r_{a\alpha}k_{ax\alpha}}}}}.
\ee
The off-diagonal elements  read
\be
C_{\alpha\beta}^{ab} =  - i{\cal N}_{a\alpha}{\cal N}_{b\beta}k_{bx\beta}\ds{\frac{z_0}{k_0}\frac{k_{a\parallel}}{k_{b\parallel}}\frac{(1-r_a)(1-r_b)}{\eps_m(k_{mzb\beta}^2-k_{mza\alpha}^2)}}\left[k_{mzb\beta}-\frac{\eps_a}{\eps_b}k_{mza\alpha}-(k_{mzb\beta}-k_{mza\alpha})\right]~~~~(\alpha,\beta >0).
\ee

Similarly, diagonal elements for two plasmons,
\be
C_{00}^{ab} = \int_{-\infty}^{+\infty} dz e_{za}h_{yb}  
={\cal N}_{a0}{\cal N}_{b0}\ds{\frac{z_0}{k_0}\frac{k_{a\parallel}}{k_{b\parallel}}}k_{bx}
\left[\ds{\frac{1}{\eps_a}\frac{1}{\kappa_{ad}+\kappa_{bd}}+\frac{1}{\eps_m}\frac{1}{\kappa_{am}+\kappa_{bm}}}\right],
\ee
determines the  normalization constant for the surface plasmon, 
\be
{\cal N}_{a0} = \ds{ \sqrt{\ds{\frac{k_0\kappa_{da}}{z_0 k_{ax}}\frac{2\eps_a\eps_m^2}{\eps_m^2-\eps_a^2}}}}.
\ee

Using the form of the electric and magnetic field of the surface plasmon, we obtain the coupling
coefficients between the surface plasmon and the plane wave in the form
\be
C_{0\beta}^{ab} = \int_{-\infty}^{+\infty} dz e_{za}H_{yb\beta} =
{\cal N}_{a0} {\cal N}_{b\beta} \ds{\frac{z_0}{k_0}\frac{k_{a\parallel}}{k_{b\parallel}}}k_{bx\beta}
\left\{\frac{1}{\eps_a}\left[\ds{\frac{r_{b\beta}}{\kappa_{ad}-ik_{bz\beta}}-\frac{1}{\kappa_{ad}+ik_{bz\beta}}}\right]+\ds{\frac{1}{\eps_m}\frac{r_{b\beta}-1}{\kappa_{am}-ik_{mzb\beta}}}\right\}.
\ee

Finally,   we obtain two sets of $N+1$ linear equations for unknown amplitudes $A$, $B$, $\oA$ and $\oB$:
\be\label{sys-1}
\begin{array}{lcl}
A-\oA &=& C^T(B-\oB)\\
B+\oB &=& C~~(A+\oA)
\end{array}
\ee
which can be rearranged into the form
\be
\mv{B}{\oA}=\mm{\bS^{bb}}{\bS^{ba}}{\bS^{ab}}{\bS^{aa}}\mv{\oB}{A}
\ee
with 
\be
\begin{array}{lclclcl}
\bS^{bb} &=& -(1+CC^T)^{-1}(1-CC^T) &~~~~~&
\bS^{ba} &=& (1+CC^T)^{-1}2C\\
\bS^{ab} &=& ~~~(1+C^TC)^{-1}2C^T &~~~~~&
\bS^{aa} &=& (1+C^TC)^{-1}(1-C^TC).
\end{array}
\ee

\acknowledgments     
 
This work was supported by  grant APVV project No.~51-003505 and  VEGA project No. 0633/09.



\end{document}